\documentclass[useAMS,usenatbib]{mn2e}
\newcommand{\OIII}{\mbox{[O III]}}
\newcommand{\NII}{\mbox{[N II]}}
\newcommand{\OI}{\mbox{[O I]}}
\newcommand{\OII}{\mbox{[O II]}}
\newcommand{\SII}{\mbox{[S II]}}
\newcommand{\HII}{\mbox{H II}}
\newcommand{\NIIHa}{\NII/H$\alpha$}
\newcommand{\SIIHa}{\SII/H$\alpha$}
\newcommand{\OIHa}{\OI/H$\alpha$}
\newcommand{\OIIIHb}{\OIII/H$\beta$}
\newcommand{\Ha}{H$\alpha$}
\newcommand{\Hb}{H$\beta$}

\usepackage{epsfig} 
\usepackage{epstopdf}
\usepackage{graphicx}
\usepackage{amsmath} 
\usepackage{hyperref}
\usepackage{verbatim}
\usepackage{tabularx}

\usepackage{breakcites}

\newcolumntype{R}{>{\centering\arraybackslash}X}

\begin{document}

\title[Starburst-AGN mixing: II. Active galaxies]{Starburst-AGN mixing: II. Optically-selected active galaxies}

\author[R. L. Davies et al.]{Rebecca L. Davies$^1$\thanks{Email:Rebecca.Davies@anu.edu.au}, Lisa J. Kewley $^{1,2}$, I-Ting Ho $^2$ and Michael A. Dopita $^{1,3}$ \\ \\
$^1$Research School of Astronomy and Astrophysics, Australian National University, Cotter Road, Weston, ACT 2611, Australia \\
$^2$Institute for Astronomy, University of Hawaii, 2680 Woodlawn Drive, Honolulu, HI 96822, U.S.A. \\
$^3$Astronomy Department, King Abdulaziz University, PO Box 80203, Jeddah, Saudi Arabia}

\maketitle

\begin{abstract}
We use 4 galaxies from the Calar Alto Legacy Integral Field Area (CALIFA) survey with clear signs of accretion onto supermassive black holes to investigate the relative contribution of star-formation and active galactic nucleus (AGN) activity to the line-emission of each galaxy as a function of radius. The combination of star-formation and AGN activity produces curved ``mixing sequences'' on standard optical diagnostic diagrams, and the fraction of emission due to AGN activity decreases smoothly with distance from the centre of the galaxy. We use the AGN activity profiles to calculate the size of the AGN narrow line regions, which have radii of \mbox{$\sim$  6.3 kpc}. We calculate the fractional contribution of the star-formation and the AGN activity to the global \Ha, \mbox{\OII\ $\lambda \lambda$ 3727,3729} and \mbox{\OIII\ $\lambda$ 5007} luminosities of each galaxy, and show that both ionization sources contribute significantly to the emission in all three lines. We use weighted combinations of stellar and AGN photoionization models to produce mixing models, which are consistent with observations for 85 percent of spaxels across the galaxies in our sample. We also present a new diagnostic for starburst-AGN mixing which requires only the \mbox{\OII}, \mbox{\OIII} and \mbox{\Hb} emission lines, and can therefore be used to calculate AGN fractions up to \mbox{z $\sim$ 0.85} in the optical and \mbox{z $\sim$ 3.5} in the near-infrared. We anticipate that this diagnostic will facilitate studies of the properties of AGN ionizing radiation fields and the relative optical contribution of star-formation and AGN activity over cosmic time.
\end{abstract}


\section{Introduction}
\label{sec:intro}

The relationship between star formation and AGN activity is one of the unsolved problems in galaxy evolution. Scaling relations between black hole mass and host galaxy properties such as stellar velocity dispersion, bulge mass and bulge luminosity strongly suggest that black holes grow simultaneously with their host galaxies \citep[e.g.][]{Kormendy95, Magorrian98, Gebhardt00, Ferrarese00, Marconi03, Haring04, Gultekin09, Kormendy09, McConnell13}. AGN activity may regulate star-formation in the host galaxy, or circumnuclear star-formation may play an important role in fuelling the accretion activity of the central supermassive black hole. However, the physical processes responsible for driving this co-evolution remain unclear. The disparity between the physical scales on which star-formation and accretion onto supermassive black holes occur has significantly mitigated attempts to develop theoretical models for such a causal connection (see \citealt{Alexander12} for a review).	 

The star-formation rates (SFRs) of AGN host galaxies are often comparable to or greater than the SFRs of inactive galaxies at the same stellar mass and redshift \citep[e.g.][]{Silverman09,Page12, Santini12, Mullaney12, Floyd13, Karouzos14}. The incidence of AGN activity increases with decreasing stellar age \citep{Silverman09}, and a total of 30-50 percent of Seyfert 2 hosts have young stellar populations \citep[e.g.][]{GonzalezDelgado01, StorchiBergmann01, CidFernandes04, GonzalezDelgado05, Sarzi07}. These results suggest that AGN activity preferentially occurs in galaxies with young stellar populations and active \HII\ regions. However, star-formation can also be quenched or suppressed through powerful galaxy-scale outflows driven by energy released during AGN accretion \citep[e.g.][]{DiMatteo05, Springel05, Bower06, Croton06, Schawinski07, Nesvadba10, DeBuhr12, Maiolino12, Karouzos14}. A large number of AGN host galaxies lie in the `Green Valley' of the galaxy colour-magnitude diagram \citep[e.g.][]{Ka03, Sanchez04, Georgakakis08}. The green colors of these galaxies suggest that star formation is on the way to being quenched as part of a transition phase from the blue cloud to the red sequence \citep{Schawinski10}. AGN activity is also commonly observed in quiescent \citep[e.g.][]{Olsen13} and post-starburst \citep[e.g.][]{Schawinski07, Cales13} galaxies. 

Recent studies have indicated that active star-formation may be a necessary condition for the onset of AGN activity, even if this star-formation is later suppressed by AGN feedback. \citet{Davies07} find evidence for a 50-100 Myr delay between the onset of star-formation and the fuelling of an AGN; suggesting that mass loss from giant stars may be an important source of low angular momentum gas to fuel the central black hole. Similarly, \citet{Wild10} find that the average black hole accretion rate increases rapidly $\sim$250 Myr after the onset of a starburst; consistent with a model in which the supermassive black hole feeds on just a small fraction of the mass shed from intermediate mass stars in the stellar bulge. The suppressed accretion efficiency during the early stages of the starburst is likely to be caused by high energy supernova feedback. 
 
Studying spatially-resolved star formation in AGN host galaxies may provide significant insight into the starburst-AGN connection. The era of Integral Field Unit (IFU) surveys for local galaxies is beginning with the commencement of programs such as the CALIFA survey (\citealt{Sanchez12}, Walcher et al. in prep), the Sydney-AAO Multi-Object Integral Field Spectrograph (SAMI) Galaxy Survey \citep{Croom12}, the Mapping Nearby Galaxies with APO (MaNGA) survey (part of SDSS-IV), the Hobby-Eberly Telescope Dark Energy Experiment (HETDEX; \citealt{Hill08}) and the Siding Spring Southern Seyfert Spectroscopic Snapshot Survey \citep[S7; e.g.][]{Dopita14}. These IFU surveys will facilitate  statistical analysis of the radial variations in star-formation histories, SFRs and the relative fraction of starburst to AGN power in AGN host galaxies. 

The \NIIHa\ versus \OIIIHb\ diagnostic diagram \citep{BPT81, Veilleux87} is commonly used to separate star-formation from AGN activity. The emission line pairs used in each ratio (\mbox{\NII\ $\lambda$ 6548} and \mbox{\Ha\ $\lambda$ 6583}, \mbox{\OIII\ $\lambda$ 5007} and \mbox{\Hb\ $\lambda$ 4861}) are very close in wavelength, making the diagnostic virtually insensitive to reddening. The \OIII/H$\beta$ ratio is sensitive to the electron temperature and ionization parameter of the line-emitting gas, and the \NIIHa\ ratio is a tracer of metallicity due to the secondary nucleosynthetic production of nitrogen at high metallicities (see \citealt{Ke13a} for further discussion). Pure star-forming galaxies and AGN host galaxies occupy two distinct sequences on the \NIIHa\ vs. \OIIIHb\ diagnostic diagram (see e.g \citealt{Ka03, Ke06, Ke13a}). The positions of \HII\ regions along the star-forming sequence are determined primarily by their chemical abundances \citep{Dopita86, Dopita00}. The hardness of the stellar ionizing radiation field increases with decreasing metallicity, perhaps due to stellar rotation or the effects of metal opacity (see e.g. \citealt{Mink09, Levesque12, Eldridge12}). The AGN sequence branches from the enriched end of the star-forming sequence and moves towards larger \NIIHa\ and \OIIIHb\ ratios as the AGN fraction and/or metallicity increases. The hard ionizing radiation field of the AGN enhances the \NIIHa\ and \OIIIHb\ ratios above their normal values for star-forming galaxies (see \citealt{Ke13a} for more details). The composite region of the AGN sequence (between the \citealt{Ke01} theoretical upper bound and the \citealt{Ka03} empirical upper bound to pure star-formation) can also be occupied by star-forming galaxies with a significant fraction of LINER-like emission \citep{Ke06} e.g. due to shock excitation or ionization by post asymptotic giant branch (pAGB) stars \citep[e.g.][]{Singh13}.

The minimum `distance' (in \NIIHa\ vs \OIIIHb\ space) of a galaxy (or spectrum) from the enriched end of the star-forming sequence can be quantified using the star-forming distance ($d_{SF}$; \citealt{Ke06}). The $d_{SF}$ metric and other variations on it have been used as relative indicators of the contribution of star formation and AGN activity to the line emission of AGN host galaxies. \citet{Wu07} defined a quantity $d_{AGN}$ which measures the distance of a galaxy from the \citet{Ke01} theoretical upper bound to pure star-formation, along lines parallel to the AGN sequence. Another common approach for calculating relative AGN fractions is to populate the composite region of the \NIIHa\ vs. \OIIIHb\ diagnostic diagram by summing observed spectra of pure star-forming and AGN-dominated galaxies with different relative weightings. The AGN fractions of observed composite spectra can then be determined by position-matching their line ratios with the line ratios of the `synthetic' composite spectra \citep[e.g.][]{Heckman04,Kauffmann09,Allen13,Richardson14}.

Unfortunately, $d_{SF}$ cannot provide an accurate estimate of the contribution of the AGN to the the total luminosity of the host galaxy, especially on th basis of a single spectrum. The metallicity and ionization parameter of the HII region and AGN narrow line region (NLR) gas and the hardness of the AGN ionizing radiation field all impact the slope and shape of the mixing curve between pure star-formation and AGN activity. Variations in the gas conditions and the nature of the radiation field between galaxies result in significant uncertainty in the absolute AGN contribution. Integral field spectroscopy resolves these issues by allowing the star forming and AGN-dominated regions to be separated spatially and spectrally.   

In Davies et al. (2014; hereafter \citeauthor{PaperI}), we presented an optical IFU study of the starburst-AGN composite galaxy NGC 7130. We found that 1) spaxels are distributed along curved mixing sequences between the pure star-forming and AGN regions on standard optical diagnostics, and 2) the dominant ionizing source is strongly dependent on distance from the centre of the galaxy. This phenomenon is known as starburst-AGN mixing. We showed that the smooth decrease in the contribution from the AGN with increasing radius can be used to calculate the relative contribution of the star-formation and AGN activity to the line-emission of each spaxel as well as the size of the AGN narrow line region.  

\begin{figure*}
\centerline{\includegraphics[scale=1,clip=true,trim=0 190 0 0]{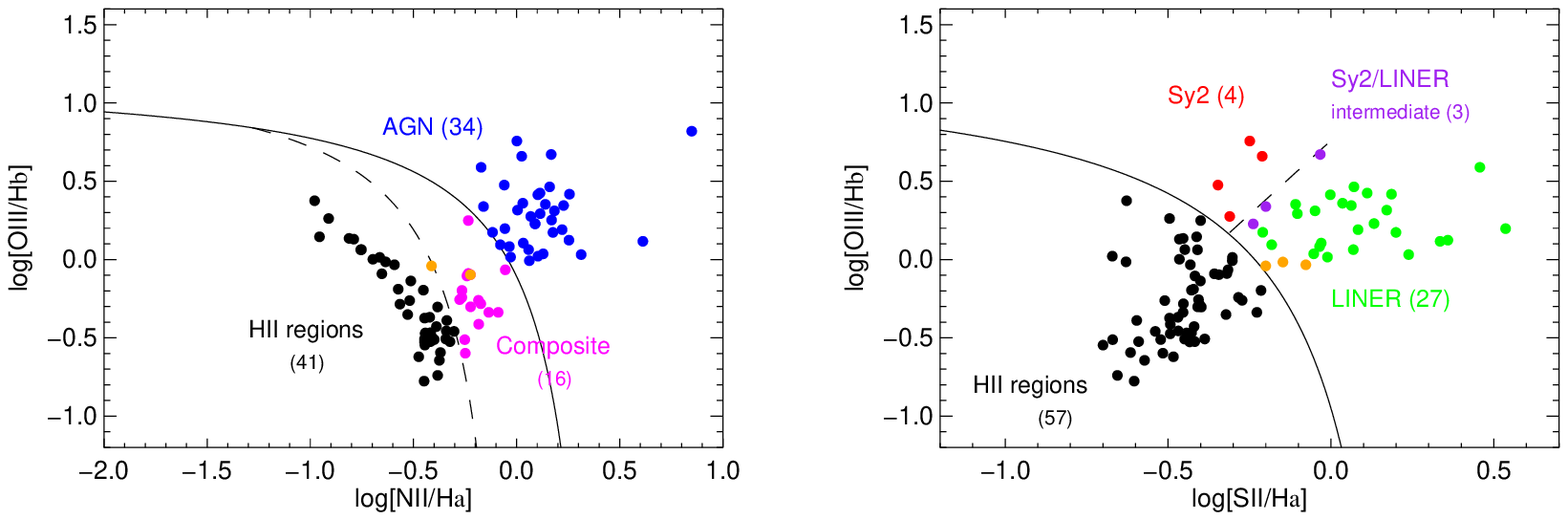}}
\caption{\NIIHa\ vs. \OIIIHb\ (left) and \SIIHa\ vs. \OIIIHb\ (right) diagnostic diagrams with line ratios extracted from \mbox{4 kpc} aperture nuclear spectra for each of the galaxies in CALIFA DR1. The galaxies are classified as either star-forming (black; 41), composite (pink; 16), Seyfert 2 (red; 4), Seyfert-LINER intermediate (purple; 3) LINER (green; 27) or ambiguous (orange; 3) according to the combined classification scheme of \citet{Ke01}, \citet{Ka03} and \citet{Ke06}.}\label{Figsample}
\end{figure*}

\begin{table*}
    \begin{tabularx}{\textwidth}{RRRRRRR}
    \hline
    Galaxy      & Recessional Velocity (km/s) (Source: \href{http://ned.ipac.caltech.edu/}{NED}) & Redshift & pc $\rm arcsec^{-1}$ & \multicolumn{2}{c}{Previous classification} & Mock-SDSS classification \\ \hline
    NGC 6394 & 8486 & 0.028 & 600 & Sy2 & \citet{Veron-Cetty06} & Sy2 \\                           
    NGC 2410 & 4861 & 0.016 & 340 & Sy2 & \citet{deVaucouleurs91} & Sy2 \\
    IC 0540 & 2035 & 0.007 & 150 & \ldots  & \ldots & Seyfert-LINER \\
    NGC 6762 & 2923 & 0.009 & 190 & \ldots  & \ldots & Seyfert-LINER 	\\ \hline
    \end{tabularx}
    \caption{Recessional velocity, redshift, spatial scale per arcsecond (based on Hubble distance) and activity-type information for the four CALIFA galaxies in our sample. We have selected AGN-dominated galaxies, rejecting LINERs because there is a high probability that they are powered by non-AGN sources \citep{Singh13}.}
    \label{table:tableA}
\end{table*}

In this paper, we extend our study of starburst-AGN mixing to active galaxies identified in the first data release of the CALIFA survey. The activity types (i.e. starburst vs. AGN) of all galaxies are determined by placing emission-line ratios extracted from mock single-fibre spectra (designed to mimic the Sloan Digital Sky Survey 3\arcsec\ fibre) on the \NIIHa\ and \SIIHa\ vs. \OIIIHb\ diagnostic diagrams (described in Section \ref{sec:sec2}). We populate the same standard optical diagnostic diagrams with emission line ratios derived from the spectrum of each spaxel in the integral field data cubes to demonstrate that the galaxies in our sample show strong evidence of starburst-AGN mixing. We calculate the fraction of line-emission due to star-formation and AGN activity in each spaxel, and use these fractions to determine the relative contributions of the star-formation and the AGN activity to the global \Ha, \OII\ and \OIII\ luminosities of the galaxies. We also calculate the size of the AGN NLR for each galaxy (Section \ref{sec:sec4}). 

We construct theoretical starburst-AGN mixing curves using a combination of stellar and AGN photoionization models, and investigate their correspondence with observed mixing sequences. We also present a new starburst-AGN mixing diagnostic which requires only the \OII, \OIII\ and \Hb\ emission lines, and discuss the potential of using this diagnostic to probe starburst-AGN mixing at high redshifts (Section \ref{sec:sec7}). We summarise the conclusions of our paper in Section \ref{sec:sec8}.

Throughout this paper we adopt cosmological parameters $H_{0} = 70.5 \, {\rm kms}^{-1}{\rm Mpc}^{-1}$, ${\Omega}_{\Lambda} = 0.73$, and $\Omega_{M}=0.27$ based on the 5-year \emph{Wilkinson Microwave Anisotropy Probe} (WMAP) results by \citet{Hinshaw09} and consistent with flat $\Lambda$ cold dark matter ($\Lambda$ CDM) cosmology. 

\section{Sample selection and emission-line fluxes}
\label{sec:sec2}

The CALIFA survey is a representative IFU survey of 600 galaxies in the local universe ($0.005 < z < 0.03$), covering a wide range of luminosities (\mbox{-23 $\la \, M_z \, \la$ -16}), colours (\mbox{0.5 $\la$ \textit{u - z} $\la$ 4}) and morphologies (as probed by the concentration index C, in 3 categories: \mbox{C $\textless$ 2.3}, \mbox{2.3 $\textless$ C $\textless$ 2.8}, \mbox{C $\textgreater$ 2.8}). The first data release (DR1) of the survey contains reduced blue (V1200, \mbox{$3650\AA \, \textless \, \lambda \, \textless \, 4840\AA$}, \mbox{R $\sim$ 1650}) and red (V500, \mbox{$3745\AA \, \textless \, \lambda \, \textless \, 7400\AA$}, \mbox{R $\sim$ 850}) data cubes for 100 galaxies. The data cubes, processed by an automatic data reduction pipeline, have a spatial size of 74\arcsec\ $\times$ 64\arcsec\ on a rectangular 1\arcsec\ grid. The point spread function (PSF) is measured from stars in the field and has a median full width at half maximum (FWHM) of 3.\arcsec7 \citep{Husemann13}. 

Emission line maps for the CALIFA galaxies are produced using the emission line fitting toolkit LZIFU written in IDL (Ho et al. in prep). LZIFU uses the  penalized pixel-fitting routine \citep[PPXF;][]{Cappellari04} to model the stellar continuum, and the Levenberg-Marquardt least-squares method to model the emission lines as Gaussians \citep[MPFIT;][]{Markwardt09}. We adopt MIUSCAT stellar models constructed by \citet{Vazdekis12}; covering 4 metallicities ([M/H] = -0.71, -0.41, 0.0, +0.22) and 50 ages (0.06 \textless\ t \textless\ 18 Gyr). We perform simple stellar population fitting with the primary aim of correcting for Balmer absorption due to old stellar populations. We note that the results of the stellar population fitting are not used to derive stellar ages or metallicities, and thus the exact choice of stellar models will have a minimal effect on our results.

We use mock 4 kpc (physical size) aperture single-fiber spectra to place all galaxies from CALIFA DR1 on the \NIIHa\ vs. \OIIIHb\ and \SIIHa\ vs. \OIIIHb\ optical diagnostic diagrams (see Figure \ref{Figsample}). \citet{Ke05} showed that activity type classifications (i.e. starburst vs. AGN) for galaxies observed as part of the Sloan Digital Sky Survey are most robust in the redshift range \mbox{0.04 $\leq$ z $\leq$ 0.1}. The lower redshift limit ensures that the aperture covers $\ga$ 20 percent of the majority of galaxies and the upper redshift limit ensures that the \SII\ doublet ($\lambda\lambda$ 6717, 6731) is not redshifted out of the spectral range of the instrument. At the mid-point of this ideal redshift range (\mbox{z = 0.07}), 3'' corresponds to \mbox{$\sim$ 4 kpc}. 

\begin{figure*}
\centerline{\includegraphics[scale=1.25,clip=true,trim=0 0 100 0]{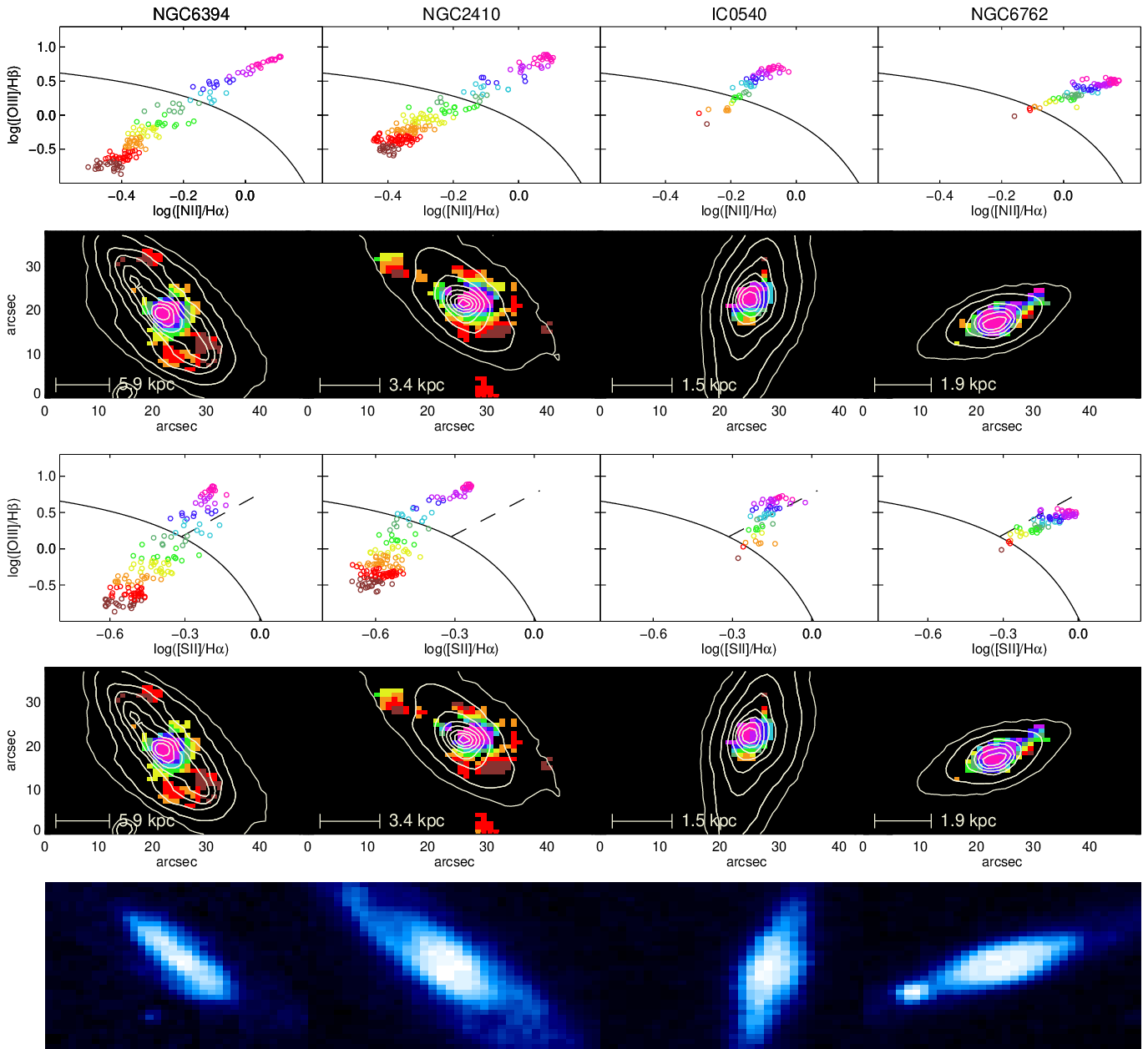}}
\caption{Row 1: \NII/H$\alpha$ vs. \OIII/H$\beta$ diagnostic diagrams with spaxels color-coded by distance along the mixing sequences. Row 2: Maps showing the locations of spaxels for which all of the BPT strong lines are detected at least the 3 sigma level, using the color-coding assigned from the mixing sequences. Rows 3 and 4 are the same as Rows 1 and 2 but using the \SIIHa\ vs. \OIIIHb\ diagnostic diagram. Bottom row: DSS (broadband) images of the galaxies indicating the optical extents of their disks. Solid lines on the diagnostic diagrams trace the theoretical upper bound to pure star-formation \citep{Ke01}. Dashed lines on the \SIIHa\ vs \OIIIHb\ diagnostic diagrams indicate the empirical division between Seyfert and LINER-like emission \citep{Ke06}. White bars on the maps show the physical scale of 10 pixels. Beige contours trace the integrated continuum emission over the wavelength range \mbox{3750$\AA$ \textless\ $\lambda$ \textless\ 7500$\AA$.} Left to right: NGC 6394, NGC 2410, IC 0540, NGC 6762. In all galaxies there is a clear relationship between the position of spaxels on the BPT diagram mixing sequences and their distance from the centre of the galaxy.}\label{FigSy2}
\end{figure*}

Each galaxy is classified as either a) pure star-forming, b) composite, c) Seyfert, d) LINER, e) Seyfert-LINER intermediate object or f) ambiguous based on the diagnostic lines of \citet{Ke01}, \citet{Ka03} and \citet{Ke06}. A full description of the combined classification scheme can be found in \citet{Ke06}. We note that our classifications may vary slightly from those of \citet{Husemann13} due to differences in the size of the line-ratio extraction aperture and the choice of classification scheme. For this study, we select a subsample of AGN-dominated objects (from categories c), d) and e)) satisfying the following conditions:

\begin{figure*}
\centerline{\includegraphics[scale=1,clip=true,trim=0 170 0 0]{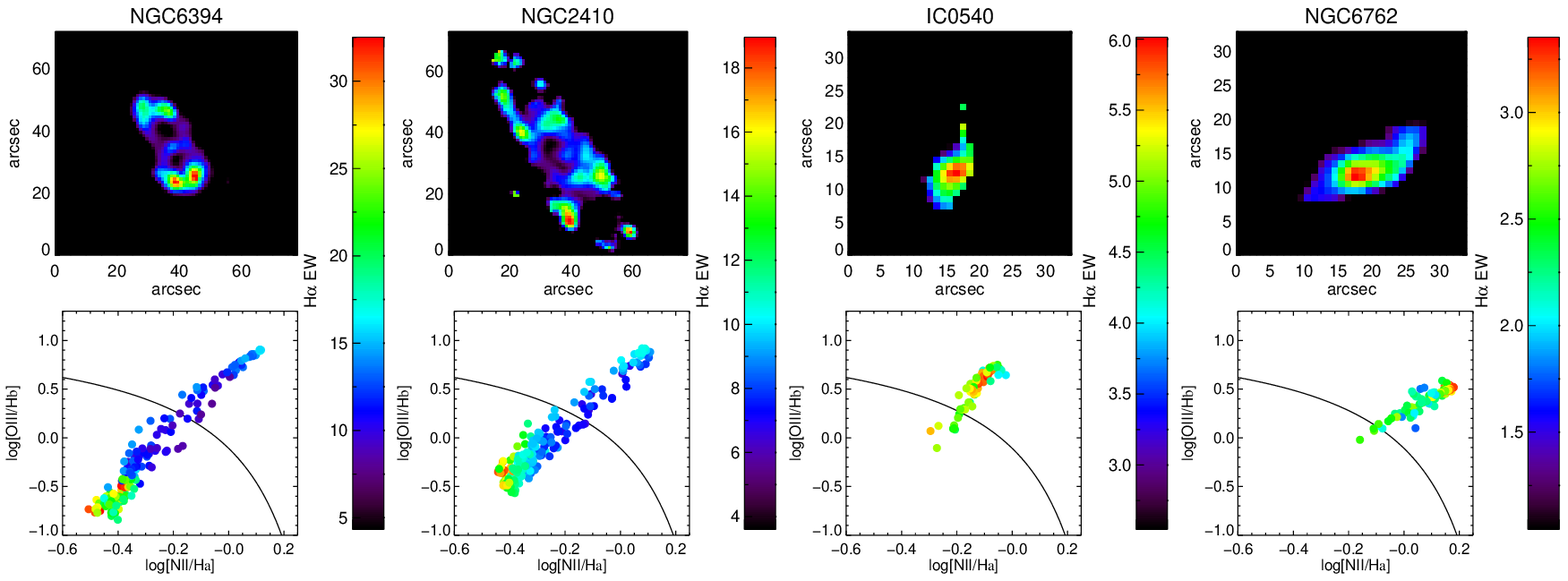}}
\caption{(Top) H$\alpha$ equivalent width ($EW_{H\alpha}$) maps covering all spaxels in which \Ha\ is detected at the 3 sigma level, and (bottom) \NIIHa\ vs. \OIIIHb\ diagnostic diagrams with spaxels color-coded by $EW_{H\alpha}$ for (left to right) NGC 6394, NGC 2410, IC 0540 and NGC 6762. More than 90 percent of spaxels in NGC 6394, NGC 2410 and IC 0540 have $EW_{H\alpha} > 3$; confirming that their emission is dominated by a combination of star-formation and AGN activity. In contrast, 90 percent of the spaxels in NGC 6762 have $EW_{H\alpha} < 3$; indicating that there is likely to be a significant contribution from post-AGB stars.}\label{FigEW}
\end{figure*}

\begin{itemize}
\item{\Ha, \Hb, \NII\ and \OIII\ must all be detected at the 3 sigma level in at least 100 spaxels. This condition aims to remove objects which have very weak emission lines and are therefore likely to be dominated by evolved stellar populations.}
\item{The \NIIHa\ and \OIIIHb\ maps must not show a strong nucleus + ring morphology. A significant fraction of the LINER galaxies in CALIFA DR1 have LINER-like emission which is spatially confined to a very small region around the nucleus, surrounded by a ring of material with very weak emission lines and then a star-forming ring at larger radii. Post-AGB stars are likely to account for the majority of the nuclear LINER-like emission \citep{Singh13}, and even if weak AGN are present in these galaxies, the sharp discontinuities in their radial emission line profiles make them unsuitable for this study.}
\end{itemize}

Our final sample consists of four galaxies: \mbox{NGC 2410} (Seyfert 2), \mbox{NGC 6394} (Seyfert 2), \mbox{IC 0540} (Seyfert-LINER intermediate) and \mbox{NGC 6762} (LINER). The recessional velocity, redshift, spatial scale per arcsecond (derived directly from the Hubble distance), previous activity-type classifications, and mock-SDSS activity-type classifications for each of the four selected galaxies are listed in Table \ref{table:tableA}.

\section{Results}
\label{sec:sec4}

\subsection{Starburst-AGN mixing}
We follow the analysis presented in \citeauthor{PaperI}. We populate the \NIIHa\ vs. \OIIIHb\ and \SIIHa\ vs. \OIIIHb\ diagnostic diagrams with emission-line ratios derived from the spectrum of each spaxel in the integral field data cube. We reject spaxels where any of the four BPT strong lines (\Ha, \Hb, \OIII\ and either \NII\ or \SII\ depending on the diagnostic) are detected at less than the 3 sigma level. The spaxels form mixing curves which lie along the SDSS AGN sequence. We split the mixing sequences into ten sections according to projected distance between the spaxels with the maximum and minimum \OIIIHb\ ratios. Row 1 of Figure \ref{FigSy2} shows the \NIIHa\ vs. \OIIIHb\ diagnostic diagrams for NGC 6394, NGC 2410, IC 0540 and NGC 6762, with spaxels color-coded by section along the mixing sequence. We use the assigned color-coding to construct maps indicating where spaxels in each section along the mixing sequence are located (Row 2). Row 3 shows the \SIIHa\ vs. \OIIIHb\ diagnostic diagrams for the same galaxies with spaxels color-coded by section along the mixing sequence, and the corresponding maps are shown in Row 4. DSS images of the galaxies are shown in the bottom panel (Row 5), indicating the optical extent of the galaxy disks.

In all galaxies, there is a strong relationship between the position of spaxels on the BPT diagram mixing sequences and their distance from the centre of the galaxy. The distinct rings seen in the maps reflect the fact that the contribution of the AGN EUV radiation field is greatest in the centre of the galaxy and decreases smoothly with radius. The stellar ionizing radiation field becomes increasingly dominant until the line emission in the extended regions of the galaxy is consistent with ionization due to pure star-formation. The maps produced from the \NIIHa\ vs. \OIIIHb\ and \SIIHa\ vs. \OIIIHb\ diagnostic diagrams are qualitatively very similar; indicating that the two diagnostics are consistent with one another. All four galaxies in our sample appear to be prototypical examples of starburst-AGN mixing.  

\subsection{H$\alpha$ equivalent widths} 
Our analysis thus far has assumed that spaxels lying in the pure star-forming region of the \NIIHa\ vs. \OIIIHb\ diagnostic diagram are dominated by \HII\ regions, and that spaxels lying in the AGN region are dominated by black hole accretion activity. Although this may appear to be a reasonable assumption, integral field spectroscopic studies have revealed that spaxels dominated by pAGB stars and/or diffuse ionized gas (DIG) can have line ratios which mimic pure star-formation or LINER-like AGN activity \citep[e.g.][]{CidFernandes10, Kehrig12, Singh13}. The \Ha\ equivalent width ($EW_{H\alpha}$) can be used to break the degeneracy between the SF/AGN and pAGB/DIG scenarios. \Ha\ is produced in the photospheres of O-B stars and can also be collisionally excited in the presence of a hard ionizing radiation field. As a result, $EW_{H\alpha}$ should be considerably higher in \HII\ regions and AGN narrow line regions ($EW_{H\alpha} \ga 6$) than in regions dominated by DIG or pAGB stars (\mbox{0.1 $< EW_{H\alpha} <$ 2.4}; \citealt{CidFernandes10, Papaderos13, Sanchez14}).

Figure \ref{FigEW} (top) shows $EW_{H\alpha}$ maps for each of the four galaxies; covering all spaxels where \Ha\ is detected at the 3 sigma level. (Note that we use positive $EW_{H\alpha}$ values to indicate \Ha\ in emission). We show \NIIHa\ vs. \OIIIHb\ diagnostic diagrams with spaxels color-coded by $EW_{H\alpha}$ in the bottom row. The color-bars are scaled to span the range of $EW_{H\alpha}$ values measured for spaxels in which all BPT strong lines are detected at the 3 sigma level (mixing sequence spaxels). 100 percent of the mixing sequence spaxels in \mbox{NGC 2410} and \mbox{NGC 6394} have $EW_{H\alpha} > 3$; confirming that the emission from these galaxies is dominated by a combination of star-formation and Seyfert 2 AGN activity. The $EW_{H\alpha}$ maps reveal complex, clumpy structures of \HII\ regions in both objects. The spaxels with the highest \Ha\ equivalent widths also appear to have the lowest AGN fractions; indicating that the \HII\ regions dominate the global production of \Ha. These findings confirm that \mbox{NGC 6394} and \mbox{NGC 2410} are prototypical examples of starburst-AGN mixing.

In IC 0540, more than 90 percent of the spaxels have $EW_{H\alpha} > 3$; indicating that the possible rate of contamination from DIG/pAGB emission is low. The \Ha\ emission is very centrally concentrated and the $EW_{H\alpha}$ decreases from the centre outwards. The spaxels with the highest $EW_{H\alpha}$ also lie strongly in the AGN region of the diagnostic diagram; suggesting that a central AGN may be responsible for the majority of the \Ha\ emission in this galaxy. Low-level emission from an underlying stellar population is likely to contribute to the optical line-emission; causing the AGN spectrum to become increasingly diluted with radius.

In contrast, 90 percent of the spaxels in NGC 6762 have $EW_{H\alpha} < 3$; indicating that pAGB stars are likely to be responsible for a significant fraction of its line emission. \citet{Kehrig12} compare the spatially resolved emission line ratios of \mbox{NGC 6762} with AGN, pAGB and fast shock models, and conclude that the pAGB scenario is most consistent with the observations. However, they note that they are unable to match the largest observed \NIIHa\ ratios without increasing the N/O ratio in the models. It is possible that mixing between a LINER AGN and pAGB stars would provide a better overall fit to the data for this galaxy, as the AGN models can easily fit the high \NIIHa\ ratios (see Section 4.3) whilst the pAGB stars provide the underlying ionization source required to dilute the AGN spectrum and produce the mixing sequence we observe. Further diagnostics and modelling are required to distinguish between these scenarios.

\subsection{Starburst-AGN fractions}
\label{subsec:AGN_fractions}

We calculate the fraction of emission due to star-formation and AGN activity for each individual spaxel in \mbox{NGC 6394} and \mbox{NGC 2410}. We cannot numerically calibrate the mixing sequences of NGC 6762 and IC 0540 to calculate AGN fractions because these galaxies do not have any spaxels which can be conclusively identified as having no contamination from AGN activity. Empirical mixing sequences are created for \mbox{NGC 6394} and \mbox{NGC 2410} by assigning the spaxels with the minimum and maximum \OIIIHb\ ratios as 0 percent and 100 percent AGN fraction respectively. The AGN fractions of all other spaxels are calculated according to their projected distance between these two extreme points. The relative error on the AGN fraction ($\Delta f_{AGN}/f_{AGN}$) of each spaxel is assumed to be 6.3 percent (see discussion in Section \ref{subsec:sec55}). 

We derive the AGN-fraction-weighted \Ha\ (\OIII) luminosity of each spaxel in NGC 6394 and NGC 2410 by multiplying the \Ha\ (\OIII) luminosity of the spaxel (\Ha$_n$, \OIII$_n$) with the corresponding (empirically calculated) AGN fraction ($f_n$). The global \Ha\ (\OIII) luminosity due to the AGN (\Ha$_{AGN}$, \OIII$_{AGN}$) is given by summing the AGN-fraction-weighted \Ha\ (\OIII) luminosities across all spaxels in each galaxy. It is then trivial to calculate the global fraction of the \Ha\ (\OIII) luminosity due to the AGN by taking the ratio of \Ha$_{AGN}$ (\OIII$_{AGN}$) to the total \Ha\ (\OIII) luminosity of the galaxy (see \citeauthor{PaperI} for a more detailed explanation of the calculations). The results are listed in Table \ref{table:table1}. 

In both galaxies, stellar photoionization dominates the global production of \Ha, whereas ionization by the hard EUV radiation field of the AGN is the dominant source of \OIII\ emission, as expected. However, the AGN is responsible for exciting at least 25 percent of the \Ha\ emission; indicating the degree to which star-formation rates calculated from \Ha\ are over-estimated in AGN host galaxies. More importantly, this fraction allows us to explore the possible contribution of AGN activity to the H$\alpha$ flux of intermediate to high redshift ``star-forming'' galaxies. Traditionally, galaxies are classified using \NIIHa\ and \OIIIHb\ ratios extracted from spectra integrated within fibres of fixed angular size (e.g. the 3\arcsec\ SDSS fibre). However, the physical covering size of such fibres increases strongly as a function of redshift. The activity type classifications of high redshift galaxies therefore reflect their globally averaged spectra rather than their nuclear spectra, which can cause actively star-forming AGN host galaxies to be classified as pure star-forming.

\begin{figure}
\centerline{\includegraphics[scale=0.55,clip=true,trim=0 150 20 0]{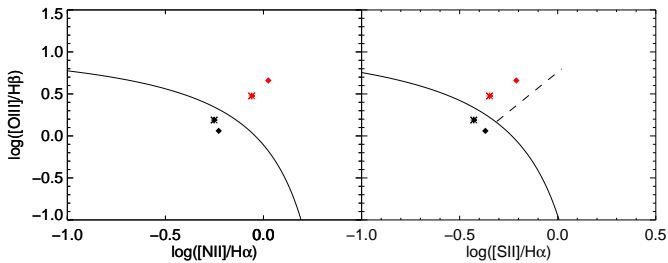}}
\caption{(Left) \NIIHa\ vs. \OIIIHb\ and (right) \SIIHa\ vs. \OIIIHb\ with line ratios extracted from (black) globally integrated spectra and (red) 4 kpc aperture nuclear spectra of (diamonds) NGC 6394 and (stars) NGC 2410. Although \mbox{NGC 6394} and \mbox{NGC 2410} are Seyfert 2 hosts, they would both be classified as star-forming galaxies if they were observed with a single fibre at intermediate to high redshift.}\label{FigBPT}
\end{figure}

To illustrate this, we have extracted \NIIHa\ and \OIIIHb\ ratios from spectra integrated across the entire field of view of the CALIFA datacubes for \mbox{NGC 6394} and \mbox{NGC 2410}. This corresponds to a physical coverage of \mbox{$\sim$ 21$\times$25 kpc} and \mbox{$\sim$ 38$\times$44 kpc} at the redshifts of \mbox{NGC 2410} and \mbox{NGC 6394} respectively. Figure \ref{FigBPT} compares the globally integrated line ratios of these galaxies (black) with line ratios extracted from \mbox{4 kpc} aperture nuclear spectra (red). Points corresponding to \mbox{NGC 2410} and \mbox{NGC 6394} have been plotted as stars and diamonds respectively. Although both \mbox{NGC 6394} and \mbox{NGC 2410} are clearly classified as Seyfert 2 hosts within a \mbox{4 kpc} nuclear aperture, their globally integrated spectra place them beneath the \citet{Ke01} theoretical upper bound to pure star-formation in both the \NIIHa\ and \SIIHa\ vs. \OIIIHb\ diagnostic diagrams. These galaxies would be classified as pure star-forming if they were observed at high redshift, but we have shown that the AGN is responsible for at least 25\% of the \Ha\ emission in both of these objects. This highlights the need to take care when calculating star-formation rates from single fibre spectra; especially at \mbox{z $\ga$ 1} where the fraction of galaxies hosting AGN is much larger than in the local universe \citep[see e.g.][]{Bundy08, Aird12}.

\begin{table*}
    \begin{tabularx}{\columnwidth}{RRRRRRRRRR}
    \cline{1-10}
    \multicolumn{2}{c}{Galaxy}         & \multicolumn{2}{c}{Star-formation (\%)}       & \multicolumn{2}{c}{AGN activity (\%)} & \multicolumn{2}{c}{NLR radius}  & \multicolumn{2}{c}{$L_{\rm NLR, \OIII}$}     \\ 
    \multicolumn{2}{c}{~} & \Ha\ & \OIII\ & \Ha\ &  \OIII\ & \multicolumn{2}{c}{(kpc)} & \multicolumn{2}{c}{~} \\ \cline{1-10}
    \multicolumn{2}{c}{NGC 6394} & 74$\pm$4 & 25$\pm$1 & 26$\pm$1 & 75$\pm$4 & \multicolumn{2}{c}{6.2$\pm$2.2} & \multicolumn{2}{c}{6.86e+40} \\
    \multicolumn{2}{c}{NGC 2410} & 63$\pm$3 & 24$\pm$1 & 37$\pm$2 & 76$\pm$4 & \multicolumn{2}{c}{6.4$\pm$1.2} & \multicolumn{2}{c}{4.04e+40} \\ \cline{1-10}
    \end{tabularx}
    \caption{Relative contribution of stellar and AGN photoionization to the global \Ha\ and \OIII\ luminosities, AGN narrow line region radii, and narrow line region \OIII\ luminosities for NGC 6394 and NGC 2410.}
    \label{table:table1}
\end{table*}

\subsection{AGN NLR radii}
We define the radius of the AGN NLR in each galaxy to be the average radius at which the AGN fraction transitions from \mbox{$\geq$ 20 percent} to \mbox{\textless\ 20 percent}. The errors on the NLR radii take into account both the dispersion in transition radii between radial vectors in each galaxy, and the smoothing of light across several pixels due to the large FWHM of the point spread functions. 

The AGN NLRs of NGC 6394 and NGC 2410 have radii of \mbox{$\sim$ 6.3 kpc} (see Table \ref{table:table1}). These radii are consistent (within the quoted errors) with the majority of Seyfert 2 NLRs which are 1-5 kpc in radius \citep{Bennert06}, but smaller than the extended narrow line regions of quasi-stellar objects which can reach out to tens of kpc from the nucleus \citep[e.g.][]{Greene11, Husemann13b}. Statistical studies of AGN NLRs have shown that their size is correlated with their total \OIII\ luminosity over three orders of magnitude in \OIII\ luminosity \citep[e.g.][]{Bennert02, Schmitt03}. The \OIII\ luminosities of the AGN NLRs of NGC 6394 and NGC 2410 (listed in Table \ref{table:table1}) are more than two orders of magnitude lower than the average \OIII\ luminosities of the quasars in the \citet{Husemann13b} sample; consistent with the significantly smaller AGN NLR radii of our objects.

We can use the largest radius at which we observe \NII, \Ha, \OIII\ and \Hb\ as a lower limit on the size of the NLR for IC 0540, which we find to be 0.47 kpc. This is small compared to the NLR radii for the Seyfert galaxies in our sample, but large compared to the average extent of \Ha\ emission in LINER galaxies (which ranges from a few tens to hundreds of parsecs; \citealt{Masegosa11}); consistent with our classification of this galaxy as a Seyfert-LINER intermediate object. The inclination of NGC 6762 to the line of sight is too large to allow for distance measurements to be accurately deprojected, preventing even a lower limit on the NLR size from being calculated. 

It is important to consider the potential impact of peculiar velocities (which can routinely be as large as a few hundred to 1000 $\rm km s^{-1}$; \citealt[e.g.][]{Lauer94,Karachentsev06, Watkins09}) on the calculated distances of the galaxies and therefore the conversion between angular and physical scales. NGC 6394 and NGC 2410 have recessional velocities of \mbox{$\sim$8500 $\rm km s^{-1}$} and \mbox{$\sim$4900 $\rm km s^{-1}$} respectively; corresponding to possible distance errors of up to $\sim$10 percent and $\sim$20 percent. These errors are quite small, indicating that any deviations between the true and calculated NLR radii will also be minimal. IC 0540 has a recessional velocity of $\sim$2000 $\rm km s^{-1}$, corresponding to a much larger potential distance error of up to 50 percent. However, increasing or decreasing the upper limit on the NLR radius of \mbox{IC 0540} by 50 percent does not change our conclusions. 

\section{Starburst-AGN mixing models}
\label{sec:sec5}

The identification of starburst-AGN mixing in all four AGN-dominated galaxies in CALIFA DR1 suggests that starburst-AGN mixing may be a common phenomenon. For this reason, we explore the possibility of using theoretical mixing curves to investigate the properties of the interstellar medium and the stellar and AGN ionizing radiation fields. If theoretical mixing sequences are able to reproduce the shape of observed mixing sequences, then they could be used to constrain the metallicity (Z) and ionization parameter (\emph{q}) of the \HII\ region and AGN NLR gas, the electron density in the AGN NLR ($n_e$) and the hardness of the AGN ionizing radiation field ($\Gamma$).

Theoretical mixing lines are constructed by linearly adding pure \HII\ region and AGN NLR spectra with different relative weightings (AGN fractions). For effective comparison with our data we construct mixing lines with AGN fractions between \mbox{0 - 100 percent} in 10 percent increments. We extract \HII\ region and AGN NLR nebular emission spectra from the grids calculated by \citet{Dopita13} and \citet{Groves04} respectively.

The shapes of the stellar and AGN ionizing radiation fields have been  characterised using the Starburst99 stellar evolutionary synthesis models of \citet{Leitherer99} and the Dusty AGN models of \citet{Groves04} respectively. These radiation fields form the input for the shock and photoionization code MAPPINGS \citep{Binette85, Sutherland93, Allen08, Dopita13}, which models the interstellar medium (ISM) and produces a simulated emission-line spectrum from the nebular gas surrounding the ionizing radiation field. 

\subsection{MAPPINGS photoionization code}

\subsubsection{Starburst99 stellar evolutionary synthesis models}
The Starburst99 stellar evolutionary synthesis models we use are described in detail in \citet{Levesque10} and \citet{Nicholls12}. These models are based on the stellar atmosphere models of \citet{Pauldrach01} and \citet{Hillier98} which consider the impact of metal opacities on the ionizing radiation field seen by the ISM. The stellar populations are evolved along the ``high'' mass-loss evolutionary tracks of \citet{Meynet94}, using a Salpeter initial mass function (IMF) \citep{Salpeter55} and zero-age instantaneous burst models which provide the best match to the SDSS \HII\ region sequence at z $\sim$ 0 \citep{Levesque10}.

\subsubsection{Dusty AGN models}
The dusty, radiation-pressure dominated AGN models we use are described in detail in \citet{Groves04}. One dimensional equilibrium photoionization models with plane-parallel geometry describe the photoionization of the AGN NLR by the ionizing radiation field from the central engine. The combination of power law AGN spectra and low electron densities ($10^2-10^4 \, \rm cm^{-3}$) in the NLR clouds produces a large number of characteristic forbidden emission lines. 

The Dusty AGN models are unique in their treatment of dust and implementation of radiation pressure physics. The models consider the absorption, scattering and polarization of light in the presence of dust. Radiation pressure from two sources is considered - the force exerted by photoionization processes, and absorption by dust. The relative significance of these two components varies with wavelength, but the latter is most significant in the far-ultraviolet (FUV) where the hydrogen opacity is effectively zero and the dust opacity is very large. By including radiation pressure physics and the impact of dust on photoionization, these models are able to account for the wide range of ionization states observed in individual narrow line regions, as well as the fast, radiatively-driven outflows observed in Seyfert galaxies (see \citealt{Groves04} for a more in-depth discussion).

The AGN models are calculated for a variety of spectral indices between \mbox{1.2 $\textless \, \Gamma \textless$ 2}. The spectral index quantifies the slope of the ionizing radiation field as a function of frequency, such that a larger $\Gamma$ correspond to a softer ionizing radiation field and a lower $\Gamma$ to a harder ionizing radiation field. 

\subsubsection{Nebular emission}
The MAPPINGS shock \& photoionization code takes an input ionizing radiation field and calculates photoionization, excitation and recombination throughout the nebula, stopping when the hydrogen gas is fully recombined. The main parameters influencing the shape of the resulting spectrum are the shape of the input ionizing radiation field and the nebular metallicity and ionization parameter. The ionization parameter \emph{q} is interpreted as the maximum velocity ionization front an ionizing radiation field is able to drive through a nebula. It has dimensions of $\rm cm \, s^{-1}$, and is related to the dimensionless ionization parameter \emph{U} by \mbox{\emph{U = q/c}}. 	 

MAPPINGS IV incorporates new atomic data and Maxwell collision strengths as well as a non-thermal ($\kappa$) electron temperature distribution (described in detail in \citealt{Nicholls12,Nicholls13}). The defining feature of this distribution is that the fraction of hot electrons can be varied using the $\kappa$ value. Setting \mbox{$\kappa$ = $\infty$} yields the Maxwell-Boltzmann distribution, and the fraction of hot electrons increases as the $\kappa$ value is decreased. Selecting \mbox{$\kappa$ = 20} produces model spectra which are consistent with observations for the vast majority of local \HII\ regions, and the resulting distribution deviates only minimally from the standard Maxwell-Boltzmann distribution \citep{Nicholls12}. 

The $\kappa$ electron temperature distribution was first introduced by \citet{Vasyliunas68} to explain the excess of hot electrons in the Earth's magnetosphere. Since then, $\kappa$ distributions have also been identified in the magnetospheres of all the gas giant planets as well as Mercury, Titan and Io (see references in \citealt{Pierrard10}) and at the interface between the solar heliosheath and the surrounding interstellar medium \citep{Livadiotis11}. These $\kappa$ distributions arise due to external energy input which prevents the system from relaxing to a classical Maxwell-Boltzmann distribution (see discussion in \citealt{Nicholls12}). Applying the $\kappa$ distribution to extragalactic \HII\ regions resolves the long standing discrepancy between electron temperatures calculated using direct, recombination line and strong line ratio methods \citep{Nicholls12}. 

\begin{figure*}
\centerline{\includegraphics[scale=1,clip=true, trim = 0 70 0 0]{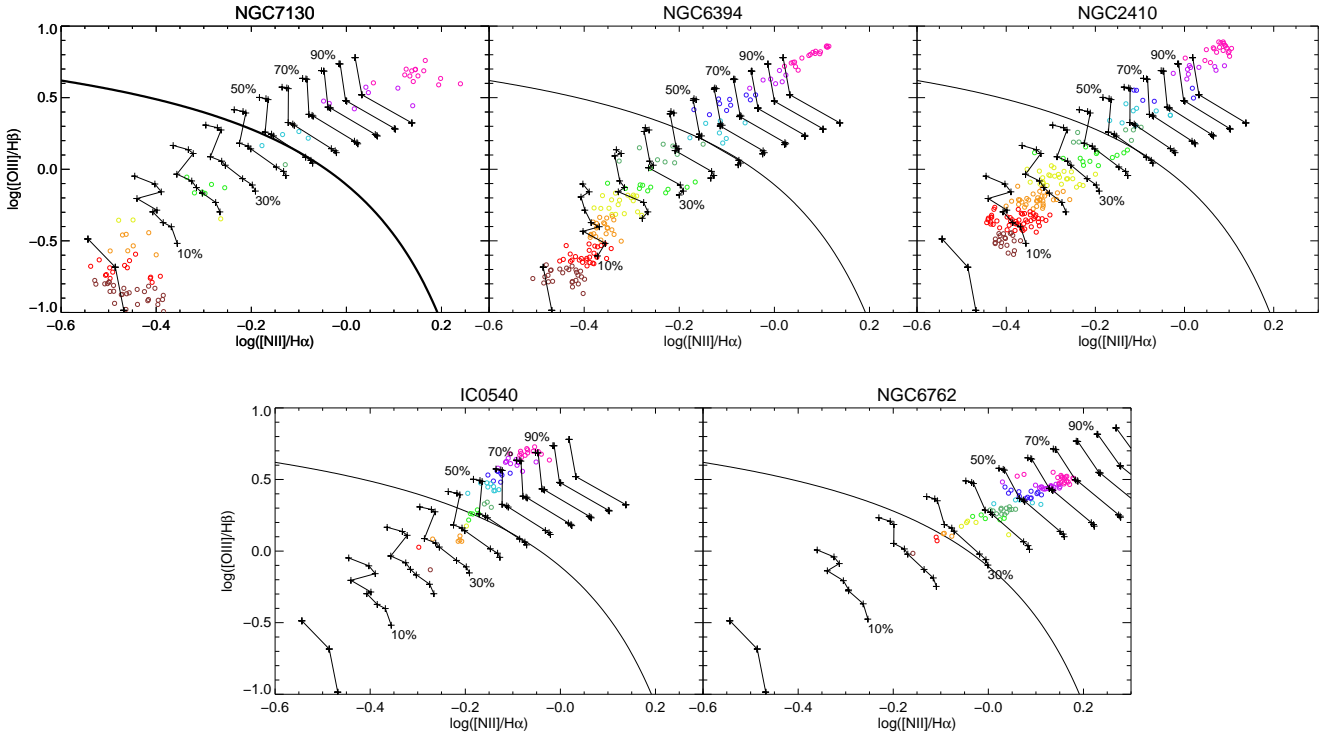}}
\caption{\NIIHa\ vs. \OIIIHb\ diagnostic diagrams for NGC 7130, NGC 6394, NGC 2410, IC 0540 and NGC 6762, with starburst-AGN mixing models overplotted as black crosses. The color-coding of spaxels is the same as in Figure 1. Mixing models are constructed by linearly adding \HII\ region and AGN NLR spectra with different relative weightings (AGN fractions). Solid lines connect models of constant AGN fraction (indicated by the percentages), spaced in 10 percent increments. We adopt models with \mbox{$\kappa$ = 20} (hot electron fraction), \mbox{$\Gamma$ = 1.7} (hardness of the AGN ionizing radiation field) and \mbox{$n_e$ = 1000 $\rm cm^{-3}$} (density of the AGN NLR clouds). The metallicity and ionization parameter values for each galaxy are selected to provide the best correspondence between the theoretical and observed mixing sequences.}
\label{Fig:models}
\end{figure*}

The grid of \HII\ region nebular emission spectra from \citet{Dopita13} is three dimensional, and covers a range of hot electron fractions \mbox{($\kappa$ = 10, 20, 50, $\infty$)}, metallicities \mbox{(0.2$Z_{\odot} \, \textless$ Z $\textless$ 5$Z_{\odot}$}; \mbox{7.99 $\textless$ 12 + log(O/H) $\textless$ 9.39)} and dimensionless ionization parameters \mbox{(-4 $\textless$ log \emph{U} $\textless$ -2)}. The grid of AGN NLR nebular emission spectra from \citet{Groves04} is four dimensional, and covers a range of electron densities \mbox{(100 $\rm cm^{-3}$ $\textless \, n_e \, \textless$ 10000 $\rm cm^{-3}$)}, AGN radiation field shapes \mbox{(1.2 $\textless \, \Gamma \textless$ 2.0)}, metallicities \mbox{(0.25$Z_{\odot} \, \textless$ Z $\textless$ 4$Z_{\odot}$}; \mbox{8.09 $\textless$ 12 + log(O/H) $\textless$ 9.29)} and dimensionless ionization parameters \mbox{(-4 $\textless$ log \emph{U} $\textless$ 0)}.

\subsection{Constraints on model parameters}
\label{subsec:sec54}

Each theoretical mixing line, produced by combining one \HII\ region spectrum with one AGN NLR spectrum, is uniquely defined by seven parameters. It is impossible to independently constrain all of these parameters using only one diagnostic diagram with four emission lines. We adopt \mbox{$\kappa$ = 20} (see discussion in Section 4.2), \mbox{$n_e$ = 1000 $\rm cm^{-3}$} (see e.g. \citealt{Stauffer82, Keel83, Phillips86}) and $\Gamma$ = 1.7 \citep[see e.g.][]{Ho93a, Groves04}, and then select the metallicity and ionization parameter values which produce the best correspondence between the theoretical and observed mixing sequence for each galaxy. It is important to note that the metallicity and ionization parameter are degenerate quantities, such that a more enriched, higher ionization model can produce the same line ratios as a more pristine, lower ionization model \citep{Ke01, Ke13a}. Detailed diagnostics employing a larger range of emission lines are required in order to discriminate between these degenerate scenarios. At present we are only concerned with the ability of the models to fit the data and not the exact parameters defining the models, and therefore discussion of the chosen model parameters is beyond the scope of this paper.

The star-formation activity of individual \HII\ regions induces small variations in the metallicity and ionization parameter of the interstellar medium. In order to account for such variations, it is possible to mix between several \HII\ region and AGN NLR spectra with similar ionization parameter values and/or metallicities. For each galaxy we extract pure \HII\ region and AGN NLR spectra at three adjacent ionization parameter values ($\Delta U = 0.3$), and mix all possible combinations of these spectra to obtain nine values for the flux in each emission line at each AGN fraction. Unfortunately, the grids are not sampled finely enough in metallicity to allow for similar treatment.

\subsection{Results}
\label{subsec:sec56}
The \NIIHa\ vs. \OIIIHb\ diagnostic diagrams for the four CALIFA galaxies and NGC 7130 are shown in Figure \ref{Fig:models}, with the chosen mixing models overplotted as black crosses. We include NGC 7130 as a known example of starburst-AGN mixing whose observed mixing sequence has not yet been compared to theoretical mixing curves (see \citeauthor{PaperI}). The chosen mixing models are consistent with observations for an average of 85 percent of spaxels across all five galaxies, and it is clear that the models are able to reproduce the shapes of the observed mixing sequences. Mixing between different ionization parameters accounts for the spread in the observed mixing sequences in the direction perpendicular to the AGN fraction vector, although the true spread appears to be smaller than possible with the models. It is interesting to note that the Seyfert-LINER intermediate objects have much less spread in their mixing sequences than the Seyfert 2 hosts. This may reflect a lack of active star formation to induce local ionization parameter variations. 

\subsection{Intrinsic scatter in AGN fractions}
\label{subsec:sec55}
The empirical AGN fraction calculations presented in Section \ref{subsec:AGN_fractions} are based on the assumption that all points along the starburst-AGN mixing sequence of a given galaxy are consistent with mixing between one star-forming spectrum and one AGN NLR spectrum. This is equivalent to assuming that all gas ionized by star-formation must have approximately the same metallicity and ionization parameter, with the same requirement for the AGN NLR gas. However, our analysis from Section \ref{subsec:sec56} indicates that there are small but significant variations in the physical conditions of the interstellar medium within galaxies. We apply our analysis from Section \ref{subsec:AGN_fractions} to theoretical mixing curves with a spread in ionization parameter, and compare the known AGN fractions of models along these curves with the AGN fractions calculated using our empirical method. This allows us to quantify the intrinsic errors on the AGN fractions calculated in Section \ref{subsec:AGN_fractions}.

\begin{figure}
\centerline{\includegraphics[scale=0.5]{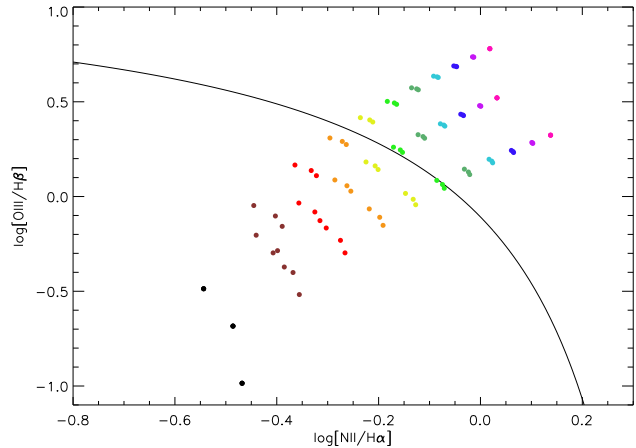}}
\caption{Theoretical starburst-AGN mixing curves produced by mixing three stellar photoionization models with three AGN photoionization models. The individual \HII\ region and AGN NLR models differ only in their ionization parameter values, which are adjacent in the model grids ($\Delta U = 0.3$). Each color represents a single AGN fraction between 0 percent (black) and 100 percent (pink), with a 10 percent increment in AGN fraction between colors.}
\label{Fig:dispersion}
\end{figure}

We show an example of a theoretical mixing sequence in Figure \ref{Fig:dispersion}, with points color-coded by AGN fraction in 10 percent increments up the sequence. Our empirical method will only recover the true AGN fractions of all model spectra when their distances along the mixing sequence are purely dependent on their AGN fractions (color in Figure \ref{Fig:dispersion}). In the pure star-forming region of the diagnostic diagram, model spectra at a given AGN fraction lie at very similar distances along the mixing sequence. However, as the contribution of the AGN increases, model spectra at a given AGN fraction become increasingly scattered along the mixing sequence, such that the relative error ($\Delta f_{AGN}/f_{AGN}$) on empirically calculated AGN fractions is \mbox{$\sim$ 6.3 percent}. This is a conservative estimate of the error given that our simulated ionization parameter variations produce mixing sequences with larger spread than observed.

Unfortunately, the models grids are too coarsely sampled in metallicity to allow for a similar investigation. Given the degeneracy between ionization parameter and metallicity in the mixing models, we expect metallicity variations to mimic ionization parameter variations - thus having minimal, if any, impact on the intrinsic errors for the calculated AGN fractions.

It is also important to consider the impact of metallicity gradients on our AGN fraction calculations. It is well established that gas-phase metallicity gradients are present in the majority of massive spiral galaxies in the local universe \citep[e.g.][]{McCall82, Zaritsky94, vanZee98}; and recent results from the CALIFA survey have confirmed and extended these findings by mapping gas-phase and stellar abundances across large numbers of local galaxies \citep[e.g.][]{Sanchez12b, Sanchez14, Sanchez-Blazquez14}. As mentioned in the introduction, the AGN sequence on the \NIIHa\ vs. \OIIIHb\ BPT diagram traces variations in both the metallicity and the ionization parameter of the line emitting gas. Therefore, metallicity gradients may impact the shapes of derived AGN fraction profiles. By constructing theoretical starburst-AGN mixing curves with different relative abundances between the starburst and AGN models (whilst holding all other parameters constant), we have determined that the presence of a smooth metallicity gradient (decreasing linearly and monotonically with radius) does not increase the dispersion in the mixing sequences. The resulting AGN fractions are consistent with those calculated from models with no metallicity gradient.

\section{Starburst-AGN mixing at high redshift}
\label{sec:sec7}

\subsection{Motivation}
We have shown that integral field spectroscopic observations of starburst-AGN mixing signatures on the \NIIHa\ vs. \OIIIHb\ diagnostic diagram provide a method for probing a) the relative contribution of star-formation and AGN activity, both spatially and as a fraction of global emission-line luminosities and fluxes, b) the physical conditions of the ISM and c) the hardness of the ionizing radiation field in AGN host galaxies. Theoretical and observational studies have indicated that the physical conditions of the ISM vary as a function of redshift and that the slope and shape of starburst-AGN mixing sequences should evolve over cosmic time \citep[e.g.][]{Ke13a,Ke13b}. Observations of starburst-AGN mixing at high redshift could provide further constraints on the evolution of the ionization parameter and the hardness of AGN ionizing radiation fields, and therefore potentially provide insight into whether (and how) the physics of AGN accretion has evolved over the last 10 billion years.

Analysis of the radial variations in the rest-frame optical line ratios of high redshift galaxies requires high resolution observations taken using NIR IFUs. Advanced adaptive optics (AO) systems have made it possible to attain PSFs with a FWHM of a few kpc for galaxies at \mbox{z $\sim$ 2}. \citet{Newman14} utilise the variations in diagnostic line ratios as a function of radius to probe the importance of star-formation, AGN activity and shock-excitation in 10 star-forming galaxies at \mbox{z $\sim$ 2} from the Spectroscopic Imaging in the Near-infrared with SINFONI (SINS) survey. \citet{Yuan11} measure the gradient in the gas-phase metal abundance of a lensed galaxy at \mbox{z = 1.49} using laser guide star (LGS) AO observations from the OSIRIS IFU on Keck II. \citet{Jorgenson14} also use OSIRIS data to construct maps of the \Ha\ and \OIII\ lines for a Ly$\alpha$-emitter at \mbox{z $\sim$ 2.2}, and \citet{Nesvadba07} present an argument for the existence of a starburst-driven wind in a sub-millimetre galaxy at \mbox{z $\sim$ 2.56} using deep NIR IFU data from SPIFFI on the VLT. These works highlight the growing potential to use the spatial variation of emission line fluxes and ratios as diagnostics for the principle ionization sources of high redshift galaxies.

Unfortunately, the redshift range over which the \NIIHa\ vs. \OIIIHb\ diagnostic diagram can be applied is fundamentally restricted by the wavelength coverage of optical and NIR IFUs. Optical IFUs such as GMOS and MUSE can observe to a maximum wavelength of \mbox{$\lambda \, \sim$ 900 nm}, and can therefore detect \NII\ and \Ha\ to \mbox{z $\sim$ 0.4} (lookback time of \mbox{$\sim$ 4.2 Gyr}), whilst the \OIII\ and \Hb\ lines will not be shifted out of the visible spectrum until \mbox{z $\ga$ 0.85} (\mbox{$\sim$ 6.6 Gyr}). In the NIR, IFUs such as NIFS, OSIRIS and SINFONI can observe to a maximum wavelength of \mbox{$\lambda \sim$ 2.5 $\mu$m}, and thus can detect \NII\ and \Ha\ to redshifts as high as \mbox{z $\sim$ 2.4} (\mbox{$\sim$ 10.9 Gyr}), whilst the \OIII\ and \Hb\ lines will not be shifted out of the NIR until \mbox{z $\ga$ 3.4} (\mbox{$\sim$ 11.8 Gyr}).

The utility of the \NIIHa\ vs. \OIIIHb\ diagnostic diagram as a probe of starburst-AGN mixing over cosmic time is also limited by the metallicity sensitivity of the \NII\ emission line. Cosmological hydrodynamical simulations have indicated that galaxies should become more enriched over cosmic time \citep[e.g.][]{Kobayashi07, Dave11}, and observations have confirmed that the mass-metallicity relation evolves towards larger metallicities as lookback time decreases \citep[e.g.][]{Erb06, Mannucci09, Zahid11, Yabe12, Zahid12b, Zahid13a, Zahid13b}. The exposure time required for robust detection of \NII\ emission in \HII\ regions therefore increases strongly as a function of redshift. Low metallicity \HII\ regions also have weak \SII\ emission, and the \OI\ emission line is much weaker than both \NII\ and \SII\ \citep[see e.g.][]{Ke06}; ruling out the \SIIHa\ and \OIHa\ vs. \OIIIHb\ diagnostic diagrams.

Starburst-AGN mixing diagnostics requiring only emission lines which lie blue-ward of \mbox{\OIII\ $\lambda$ 5007} and are strong at low metallicity would significantly increase the redshift range over which starburst-AGN mixing can be probed. The strong, low ionization \mbox{\OII\ $\lambda \lambda$ 3727,3729} doublet is often used as a star-formation rate indicator in the absence of \Ha\ measurements \citep[e.g.][]{Gallagher89, Lilly96, Kennicutt98, RosaGonzalez02, Ke04}. The \OII\ emission lines are excited by hot electrons and therefore are strongest in low metallicity \HII\ regions which do not have significant metal cooling \citep{KD02}. Here, we investigate whether the \OIII/\OII\ line ratio can be used in conjunction with the \OIIIHb\ ratio to probe starburst-AGN mixing. 

\begin{figure*}
\centerline{\includegraphics[scale=1,clip=true,trim=0 165 0 0]{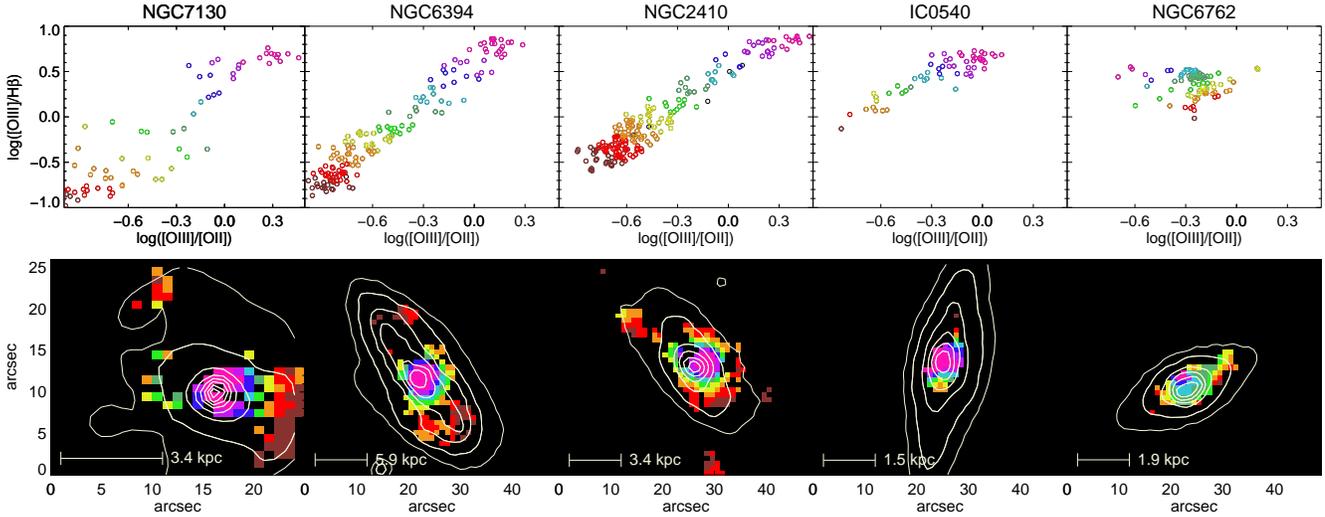}}
\caption{(Top) $\rm [O III]_{corr}$/$\rm [O II]_{corr}$ vs. \OIII/H$\beta$ diagnostic diagrams with spaxels color-coded by distance along the mixing sequence, and (bottom) maps indicating where spaxels in each section of the mixing sequence are located, using the color-coding assigned from the mixing sequences. Beige contours trace the integrated continuum emission over the wavelength range \mbox{3750$\AA$ \textless\ $\lambda$ \textless\ 7500$\AA$.} Left to right: NGC 7130, NGC 6394, NGC 2410, IC 0540, NGC 6762. Bars indicate the physical scale of 10 pixels on the maps. In all galaxies there is a clear relationship between the position of spaxels on the mixing sequence and their distance from the centre of the galaxy - similar to what is found for the \NIIHa\ and \SIIHa\ vs. \OIIIHb\ diagnostic diagrams.}\label{Fig:OIII_OII}
\end{figure*}

\subsection{Extinction correction}
The combination of the \OIII/\OII\ and \OIIIHb\ line ratios was first proposed as a diagnostic for the power sources of galaxies by \citet{BPT81}. However, the wavelength separation of the \mbox{\OIII\ $\lambda$ 5007} and  \mbox{\OII\ $\lambda \lambda$ 3727,3729} lines makes the \OIII/\OII\ ratio very sensitive to interstellar extinction, and therefore the reddening-insensitive \NIIHa, \SIIHa\ and \OIHa\ vs. \OIIIHb\ diagrams have become the `universal' diagnostics for the power sources of galaxies. We correct the \OIII, \OII\ and \Hb\ emission line fluxes for interstellar extinction by comparing the intrinsic \Ha/\Hb\ ratio (Balmer decrement) with its measured value in each individual spaxel (assuming the \citealt{Cardelli89} extinction law). The \Ha\ and \Hb\ emission lines originate from recombination to the \mbox{n = 2} level of hydrogen, and therefore the intrinsic \Ha/\Hb\ ratio can be accurately calculated using the relative lifetimes of the \Ha\ (\mbox{3 $\rightarrow$ 2}) and \Hb\ (\mbox{4 $\rightarrow$ 2}) transitions. Assuming Case B recombination, the intrinsic Balmer decrement is $\sim$2.86 \citep{Osterbrock89}. Collisional excitation enhances the strength of \Ha\ relative to \Hb; causing the Balmer decrement to rise to $\sim$3.1 in the presence of an AGN. 

We assume that the intrinsic Balmer decrement is 2.86 in all spaxels. The validity of this assumption varies with AGN fraction; but in the worst case scenario (for spaxels with 100 percent AGN fraction in the nuclear regions of galaxies), the corrected \Ha\ fluxes will be over-estimated by a factor of 1.08. Deviations in the local electron temperature and density can also introduce up to 5\% uncertainty on the intrinsic Balmer decrement \citep{Osterbrock89}. Taking both effects into account, there will be a maximum of \mbox{0.06 dex} difference between the corrected and intrinsic $\log$(\Ha/\Hb) ratios. We expect similar errors on the derived $\log$($\rm [O III]_{corr}$/$\rm [O II]_{corr}$) ratios, given that \OIII\ and \OII\ lie only \mbox{1000 \AA} blueward of \Hb\ and \Ha\ and have a smaller wavelength separation (1280 \AA\ compared to 1722 \AA). Errors of this size are considerably smaller than the intrinsic spread in line ratios at a single AGN fraction (see Section \ref{subsec:sec55}) and are therefore negligible.

If the Balmer decrement is used to correct for interstellar extinction, the suite of lines required for our proposed starburst-AGN mixing diagnostic consists of \Ha, \Hb, \OIII\ and \OII. This removes the need to observe lines which are weak at low metallicity, but the wavelength of the \Ha\ line still strongly restricts the maximum redshift to which the diagnostic can be applied. In order to alleviate this issue, the $EW_{[O III]}$/$EW_{[O II]}$ equivalent width ratio may be used as a partially reddening-insensitive proxy for the \OIII/\OII\ line flux ratio. The dependence of equivalent width on stellar continuum strength automatically corrects the ratio for any reddening affecting the stars, making the equivalent width ratio superior over the line flux ratio when reddening corrections are not available \citep{Kobulnicky03, Kobulnicky04}. The equivalent width ratio is related to the reddening-corrected line flux ratio by some normalization constant $\alpha$;  i.e. \mbox{$\rm [O III]_{corr}/[O II]_{corr}$ = $\rm \alpha (EW_{[O III]}$/$EW_{[O II]})$}. The size of $\alpha$ is determined by both the ratio between the reddened and intrinsic line flux ratios (which depends on the underlying stellar population) as well as the difference in the amount of attenuation experienced by the ionized gas and the stars. \citet{Kobulnicky03} show that, assuming an average galaxy stellar spectrum dominated by A-type stars and G \& K giants, $\alpha \sim 0.84 \pm 0.3$. The average value of log($\alpha$) therefore falls short of unity by less than 0.1 dex. Use of the equivalent width ratio also alleviates the need for flux calibration, which is notoriously difficult for NIR observations. Any calibration affects the strength of the emission lines and the stellar continuum equally; leaving the equivalent width measurements unchanged \citep{Kobulnicky03}.

The extinction along the line of sight can also be calculated using alternate techniques, including higher order Balmer line ratios (e.g. \mbox{H$\gamma \, \lambda$ 4341}/\Hb), photometric colors or the ratio of the total infrared (TIR) to UV luminosity \citep[e.g.][]{Buat92, Meurer99}.

\subsection{AGN fractions using \OIII/\OII\ and \OIIIHb}
\label{subsec:sec62}
We show the $\rm [O III]_{corr}$/$\rm [O II]_{corr}$ ratio compared to the \OIIIHb\ ratio for the five galaxies in our sample (again including NGC 7130) in the top row of Figure \ref{Fig:OIII_OII}. We observe a mixing sequence for the $\rm [O III]_{corr}$/$\rm [O II]_{corr}$ vs. \OIIIHb\ diagram, similar to the \NIIHa\ vs. \OIIIHb\ diagnostic diagram. We color-code spaxels according to projected distance along the line between the spaxels with the minimum and maximum \OIIIHb\ ratios, and use the same color-coding to construct maps indicating where spaxels in each section of the mixing sequence are located (as in Figure \ref{FigSy2}). The position of spaxels on the $\rm [O III]_{corr}$/$\rm [O II]_{corr}$ vs. \OIIIHb\ mixing sequence is directly related to their distance from the centre of the galaxy. 

It is interesting to note that \mbox{NGC 6762} appears to have distinctly different properties from the rest of the galaxies in our sample. No mixing sequence is visible, and there is no clear correlation between the \OIII/\OII\ and \OIIIHb\ ratios. \citet{BPT81} showed that \HII\ regions fall along a tight sequence on the \OIII/\OII\ vs. \OIIIHb\ diagram, with a strong positive correlation between the two ratios. However, objects ionized by harder ionizing radiation fields, such as planetary nebulae, AGN activity or shocks, do not lie along this correlation. The absence of a correlation between the \NIIHa\ and \OIIIHb\ ratios in \mbox{NGC 6762} suggests that young stars are not a significant source of ionization; consistent with the low $EW_{H\alpha}$ values observed in this object. This finding highlights the possibility to use the \OIII/\OII\ vs. \OIIIHb\ diagram as an additional tool to distinguish between galaxies which are ionized by a combination of active star-formation and AGN activity and interlopers which are dominated by older stellar populations.

\begin{figure}
\centerline{\includegraphics[scale=0.5]{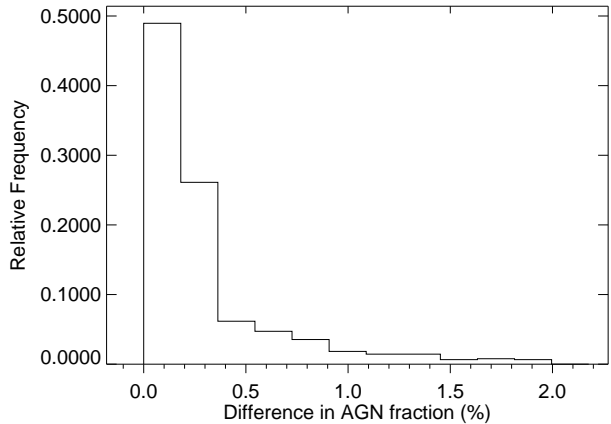}}
\caption{Difference in AGN fractions derived from the \NIIHa\ vs. \OIIIHb\ and \OIII/\OII\ vs. \OIIIHb\ diagnostic diagrams, taking into account all spaxels with calculated AGN fractions from NGC 7130, NGC 6394 and NGC 2410. The difference is less than 2 percent for 96 percent of spaxels, indicating very good quantitative correspondence between the diagnostics.}
\label{Fig:hist}
\end{figure}

We apply the technique described in Section \ref{subsec:AGN_fractions} to the  $\rm [O III]_{corr}$/$\rm [O II]_{corr}$ vs. \OIIIHb\ mixing sequences and thereby derive a second estimate for the AGN fraction of each spaxel in NGC 7130, NGC 6394 and NGC 2410. We quantify the level of consistency between the \OIII/\OII\ vs. \OIIIHb\ and \NIIHa\ vs. \OIIIHb\ diagnostics by comparing the AGN fraction estimates for each spaxel individually. The distribution of the difference in AGN fractions between the two diagnostics is shown in Figure \ref{Fig:hist}. The diagnostics agree to within 0.5 percent in 80 percent of spaxels, 1 percent in 91 percent of spaxels and 2 percent in 96 percent of spaxels, indicating remarkably good correspondence. This confirms that the \OIII/\OII\ vs. \OIIIHb\ diagnostic can be used as a tracer of starburst-AGN mixing at high redshifts, provided dust attenuation has been accounted for (e.g. using the techniques described in the previous section).

\subsection{Enhancement of \OII\ emission due to AGN activity}
The \OII\ emission line is commonly used as a SFR indicator in AGN host galaxies, under the assumption that this line is only weakly enhanced in the presence of an AGN \citep{Ferland86, Ho93a, Ho93b, Ho05, Kim06, Silverman09}. In order to test this assumption, we calculate the relative contribution of the star-formation and the AGN activity to the global \OII\ luminosities of NGC 7130, NGC 6394 and NGC 2410, using the technique outlined in Section \ref{subsec:AGN_fractions}. The results can be found in Table \ref{table:table2}. We have assumed the relative error on AGN fractions calculated from the \OIII/\OII\ vs. \OIIIHb\ diagnostic to be the same as the error on those calculated from the \NIIHa\ vs. \OIIIHb\ diagnostic. 

The AGN is responsible for exciting approximately 40 percent of the global \OII\ emission in all three galaxies, which is 10 percent higher than the average contribution of the AGN to the \Ha\ emission in the same galaxies. We therefore do not recommend the use of uncorrected global \OII\ luminosities to estimate the star formation rate of a galaxy.  Instead, we recommend the use of integral field spectroscopy to remove the AGN contribution before the star formation rate is calculated.

\begin{table}
    \begin{tabularx}{\columnwidth}{RRR}
    \hline
    Galaxy        & Star-formation (\%)    & AGN (\%) \\ \hline
    NGC 7130 & 60 $\pm$ 3 & 40 $\pm$ 2  \\
    NGC 6394 & 56 $\pm$ 3 & 44 $\pm$ 2 \\
    NGC 2410 & 56 $\pm$ 3 & 44 $\pm$ 2  \\ \hline
    \end{tabularx}
	\caption{Relative contribution of star-formation and AGN activity to the global \OII\ luminosities of NGC 7130, NGC 6394 and NGC 2410.}
    \label{table:table2}
\end{table}

\section{Conclusions}
\label{sec:sec8}

We have analysed the relationship between star-formation and AGN activity as a function of radius in four AGN host galaxies using optical IFU data from the CALIFA survey. We construct maps of the spaxels in each galaxy for which all of the BPT strong lines are detected at the 3 sigma level, color-coded by distance of the spaxels along the starburst-AGN mixing sequence on the \NIIHa\ vs. \OIIIHb\ diagnostic diagram. We observe clear rings of gas ionized with decreasing contributions from the AGN as radius increases.

We have numerically calibrated the starburst-AGN mixing sequences to calculate the fractional contribution of the star-formation and the AGN activity to the line-emission of each spaxel in each galaxy. We find that the star-formation and the AGN are each responsible for at least 25 percent of the global \Ha, \OII\ and \OIII\ luminosities in all galaxies - highlighting the need to correct for any AGN contribution when calculating star-formation rates using \Ha\ or \OII. 

We have constructed theoretical starburst-AGN mixing lines using weighted combinations of HII region and AGN narrow line region nebular emission spectra. We have shown that these models are consistent with observations for an average of 85 percent of spaxels across all galaxies, and are able to reproduce the slope and shape of observed mixing sequences. This agreement highlights the potential to use starburst-AGN mixing models to constrain the ionization parameter, metallicity and hardness of the AGN ionizing radiation field in AGN host galaxies.

We have shown that the \OIII/\OII\ vs. \OIIIHb\ diagram provides an alternate diagnostic to the \NIIHa\ vs. \OIIIHb\ diagram for calculating AGN fractions. We anticipate that the \OIII/\OII\ vs. \OIIIHb\ diagram will be useful for separating starburst and AGN activity to high redshifts (z $\ga$ 2.5).

Many IFU surveys are currently underway (including SAMI, CALIFA, S7, MaNGA and HETDEX) and many more integral field spectrographs will be coming online in the near future (including the Multi Unit Spectroscopic Explorer (MUSE) for the VLT \citep{Bacon10}, the Keck Cosmic Web Imager (KCWI; \citealt{Martin11}) and the Giant Magellan Telescope Integral Field Spectrograph (GMTIFS; \citealt{McGregor12})) to survey galaxies both locally and at high redshift. These instruments will provide statistically significant samples of IFU data to investigate the relationship between star-formation and AGN activity across cosmic time. The techniques demonstrated in this paper provide just one of the first stepping stones towards using these unique datasets to gain a greater understanding of how black holes and their host galaxies have evolved over the history of the universe.

\section{Acknowledgements}

The authors would like to thank the anonymous referee whose comments and suggestions facilitated significant improvement of this paper. Kewley and Dopita acknowledge the support of the Australian Research Council (ARC) through Discovery project DP130103925. Dopita acknowledges financial support received under the KAU HiCi program. This research has made use of the data provided by the Calar Alto Legacy Integral Field Area (CALIFA) survey (\href{http://califa.caha.es/}{http://califa.caha.es/}). Based on observations collected at the Centro Astron\'omico Hispano Alem\'an (CAHA) at Calar Alto, operated jointly by the Max-Planck-Institut f\"ur Astronomie and the Instituto de Astrofisica de Andalucia (CSIC). This research has made use of the NASA/IPAC Extragalactic Database (NED) which is operated by the Jet Propulsion Laboratory, California Institute of Technology, under contract with the National Aeronautics and Space Administration. We also acknowledge the usage of the HyperLeda Database \href{http://leda.univ-lyon1.fr}{(http://leda.univ-lyon1.fr)} and Aladin Sky Atlas.

\end{document}